\documentclass{ws-rv10x7}
\usepackage{ws-rv-thm}
\usepackage{mathtools,xfrac}
\usepackage[square]{ws-rv-van}
\usepackage[pdfpagelabels=false,colorlinks=true,allcolors=black]{hyperref}
\renewcommand\thesection{\thechapter.\arabic{section}}
\renewcommand\theequation{\thechapter.\arabic{equation}}
\renewcommand\thefigure{\thechapter.\arabic{figure}}
\usepackage{perpage}\MakePerPage{footnote}

\makeindex

\newcommand{\be}{\begin{equation}}
\newcommand{\ee}{\end{equation}}
\newcommand{\bea}{\begin{eqnarray}}
\newcommand{\eea}{\end{eqnarray}}

\renewcommand{\>}{\right\rangle}

\newcommand{\nn}{\nonumber }

\begin{document}

\setcounter{chapter}{17} 

\chapter[Quantum glasses]{Quantum Glasses}\label{ch1}
\author[L.F.Cugliandolo and M.M\"uller]{Leticia F. Cugliandolo$^{1,2}$ and Markus M\"uller$^3$}
\address{$^1$Sorbonne Universit\'e, Laboratoire de Physique Th\'eorique et Hautes Energies, CNRS
UMR 7589, 4 Place Jussieu, 75252 Paris 05, France
\\
$^2$Institut Universitaire de France, 1 rue Descartes, 75231 Paris 05, France
\\
$^3$Laboratory for Theoretical and Computational Physics, \\
Paul Scherrer Institute, Villigen CH-5232, Switzerland}

\section{Introduction}
\label{intro}

The interplay of disorder, interactions and quantum mechanics leads to a host of interesting phenomena. 
A particularly intriguing aspect is the breakdown of ergodicity in a way that differs significantly from that of standard spontaneous symmetry breaking. 
This can arise as a consequence of quenched disorder, as in spin glasses, or structural disorder, 
think of amorphous density order. Both quenched and self-generated disorder 
entail a clustering in phase space associated with glass attributes like memory effects and extremely slow relaxations~\cite{Bouchaud,BerthierBiroli,SchmalianWolynes03}.
The replica method allows one to access the phase space structure,
and  replica symmetry breaking (RSB) 
signals the existence of a large number of metastable states separated by 
substantial free energy barriers. Even though barriers may be finite  in low dimensions, preventing a genuine thermodynamic 
glass transition, the replica method has proven useful for these cases as well.

As a result of frustration and the competition between many nearly degenerate ordering patterns, 
the glassy order is often rather soft and gives rise to an unusually high density of low energy excitations.
In or close to a quantum glass phase those may have a strong influence on electronic transport and have often been invoked as candidates to explain the non-Fermi liquid behavior of the resistivity of complex materials, such as high temperature superconductors (cuprates), heavy fermion systems, but also systems close to a Mott-Anderson  metal-insulator transition. 

{Apart from this ``glassy" route,
very slow dynamics and non-ergodicity in quantum systems can be due to other mechanisms too, especially in low dimensions where the effects of disorder
are very strong.} 
On the one hand, under a real space 
renormalization group treatment, disordered quantum spin chains typically flow to fixed points characterized by infinite
 randomness~\cite{DSFisher1992}, with extended regions of parameter space where Griffiths--McCoy singularities are very important~\cite{RiegerYoung,IgloiMonthus2005}. Thereby the low frequency response is dominated by nearly isolated degrees of freedom only weakly coupled to others further away. 

On the other hand, non-interacting quantum particles in one dimension subject to a disordered potential Anderson localize and stop diffusing on the scale of the elastic mean free path~\cite{Anderson1958,MottTwose1961}. 
{Its interacting counterpart,} many-body localization (MBL)~\cite{Basko2006,Nandkinshore,Alet,Abanin2019},
relies on the discreteness of non-interacting excitations and the rareness of resonant interactions, which then fail to induce diffusion of energy. Thus, MBL can only occur in isolation, with no coupling to the continuum of a bath which would reinstate transport and ergodicity.  In contrast, the above discussed glass phases  
are stable against a bath, as they rely on the rugged structure of the free-energy landscape. 
A further difference between quantum glasses and MBL systems is that the latter are highly susceptible to local inclusions where disorder is weak. Those act as sources of dynamical chaos, and it is believed that only in low-dimensional, discrete lattices such nucleation centers of local baths do not destroy MBL~\cite{Imbrie,HuveneersDeRoeck,Schiulaz} (although the existence of MBL even in 1d has been questioned recently~\cite{Prosen2020, Sels2021, Kiefer2021}).
In contrast, glasses, like standard symmetry broken phases, are stabilized in high dimensions (and RSB is certain to occur above a $d_c$~\cite{DeDominicis}).
In summary, while both quantum glassiness and MBL entail non-ergodicity and impede full thermalization,
they do so for fundamentally different reasons.

%other sources of ergodicity breaking
Other obstructions to full ergodicity exist, {\it e.g.}, integrable systems with an  extensive number of conserved quantities~\cite{Calabrese},  systems hosting many-body scars (a set of eigenstates that violate the eigenstate thermalization hypothesis)~\cite{Turner2018} or cases with fragmented Hilbert spaces~\cite{Moudgalya_2022}. However, none of these is robust against generic perturbations of the Hamiltonian, as even modest ones 
 reinstate equilibrating dynamics rather quickly. We do not consider them as quantum glassy 
 and we will not discuss them here. 
 A further way to kill ergodicity is the sufficiently strong coupling to an Ohmic bath of a single quantum degree of freedom,  which can get dressed with many bath modes, so as to acquire an infinite mass, and its dynamics freeze out~\cite{MFisher,MVojta}. 
 We shall not discuss this effect of non-glassy ``local ergodicity'' breaking either, except for noting that 
the coupling of quantum systems to a bath usually results in the reduction of fluctuations, and thus extends 
the regime of stability of the glass~\cite{bath-spherical,bath-SU2}.

% high dim disorder + frustration = glass in the sense of amporphous order, RSB etc.  - stability/phase diagram
{In this review  we focus on quantum glasses that arise in high dimensions,} where disorder is so relevant
as to produce amorphous order and phase space clusterization,  
but does not flow to infinite randomness.  
This structure withstands both thermal {\em and} quantum fluctuations, 
which soften the energy landscape and eventually melt the glass. For certain types of glasses the quantum melting  differs   
drastically from the thermal route. 

Since this chapter is part of a book devoted to replica symmetry breaking, we focus the 
discussion on equilibrium aspects and we structure it as follows.
In Sec.~\ref{sec:dynamics} we overview some general aspects of quantum glassy dynamics. Section~\ref{replicatreatment} reminds the gist of the replica treatment of mean field glasses.
In Sec. \ref{sec:classification}  we classify  different universality classes of (mean field) glasses according to the type of interactions and environments, and discuss their phenomenology, in particular the nature of their low energy collective modes which we obtain both from Landau theory and an effective potential approach in Sec.~\ref{sec:Landau}.
In Sec.~\ref{sec:qpspin} we discuss whether an isolated quantum glass can explore its clusterized phase space by tunneling or stays localized.
The interplay between the glass and localization routes to non-equilibrium is a multi-faceted topic
briefly reviewed in Sec.~\ref{sec:interplay}. 
Section~\ref{sec:experimentalsystem}  presents some experimental glassy materials.
Finally, in Sec.~\ref{sec:SYK} we survey  fermionic models closely connected to the paramagnetic phase of Heisenberg spin glasses, {\it a.k.a.} the Sachdev-Ye-Kitaev (SYK) model, that may 
fertilize the study of spin glasses via interesting dynamical analogies.

%dynamics: long= classical; short : reflects quantum nature,
\section{Dynamics in quantum glassy phases }
\label{sec:dynamics}

{A particularly interesting aspect of quantum glasses is their dynamics, both in and out of equilibrium. We broadly survey them here and discuss selected questions in more detail in later sections. }
%We give here a broad survey of interesting aspects of the quantum dynamics in a glass phase.  Some topics will be discussed in more depth later.

\subsection{Short-time dynamics}

Distinct quantum aspects of glasses appear on microscopic time scales.
The high-frequency dynamics depends crucially on  the symmetries of the order parameter and the nature of the quantum fluctuations. 
In the low-frequency limit, instead, the excitations feel the softness of the glassy landscape, %As we will see, 
%of infinitely long ranged interactions, 
which, in the mean-field limit, exhibits surprisingly universal spectral features whose frequency scaling merely depends on the glass being metallic or insulating. 
We will review the salient results in Secs.~\ref{sec:classification}, \ref{sec:Landau} and \ref{sec:SYK}.
%on  the symmetries of the order parameter and the nature of the quantum fluctuations. 

\subsection{Moderately long-time dynamics}

The intermediate to long-time dynamics of quantum glassy systems are essentially identical to those of  their classical counterparts. {Indeed, one expects that 
 on time scales $t \gg \hbar/T$ all degrees of freedom decohere and any quantum effects are washed out.} After a quench from the disordered phase, 
the glass hovers over saddles in the free energy landscape and slowly relaxes its energy without ever getting deeply trapped into rare deep minima, in a manner dubbed {\it weak ergodicity breaking}~\cite{Culo,Culo1,Bicu,KennettChamon,BiroliParcollet02}. The response slows down increasingly with time (the glass ``ages''). In solvable mean field models, the fluctuation-dissipation relation evaluated on those long time scales takes the classical form, albeit with an effective temperature $T_{\rm eff}$ that exceeds the bath temperature
(even at $T=0$).
%, and increases with time. 
The latter is formally related to Parisi's order parameter  $Q(x)$ as in the classical limit~\cite{Bouchaud}.
 Moreover, the  equations governing the slow evolution of the response and correlation functions are invariant under time reparametrizations, $t\to f(t)$~\cite{ChamonCugliandolo}.  
 Such an invariance also occurs  in the large $M$ limit of the $SU(M) $
Heisenberg and the SYK models. Recent progress in this direction will be reviewed in Sec.~\ref{sec:SYK}. 
Finally, as in the classical limit, one can tune the initial state to be within one of the valleys in the free-energy landscape 
and show that the subsequent dynamics will remain confined to it (at least for times exponentially large in system size). 

A short summary of the real-time dynamics of 
quantum disordered systems on these time scales can be found in Ref.~\cite{Cugliandolo04}. 
We will not extend this discussion further 
here, our aim being to focus mainly on equilibrium properties. 

\subsection{Extremely long-time dynamics
%- Tunneling and quantum annealing
}

%\subsubsection{Quantum annealing}
On extremely long time scales, crossing over or tunneling under barriers between metastable valleys
should be possible.
In mean-field models barriers scale with system size 
%these times typically scale exponentially with the system size. 
%The question then arises as to whether quantum tunnelling may speed up the time the system needs to explore low energy valleys  - as compared to 
and  thermal activation over them  is exponentially slow. 
Since the latter  is the root cause for the failure of simulated annealing in finding the ground state of 
NP-hard  problems, 
there was hope that quantum annealing could achieve a significant speed-up~\cite{Nishimori}. 
The idea was to
% to find the ground state of a complex glassy system  by 
initialize the system under strong quantum fluctuations ({\it e.g.} a transverse field) which
% destroy the glassy order and 
induce a simple ground state. Then, by invoking the adiabatic theorem, one should reach 
the classical ground state upon turning off the fluctuations  sufficiently slowly. 
The caveat is that exponentially slow switch-off rates 
are required to follow the ground state while it delicately 
hybridizes distant low-energy valleys in phase space. Still, one may  ask which of the  exponentially long times (barrier crossing or adiabatic ground state preparation) is more efficient to explore complex 
landscapes~\cite{Baldwin18}, {an idea that was explored experimentally in quantum Ising spin glasses ~\cite{Brook}, 
cf.~Sec.~\ref{sec:QIsingspinglass}}.   
% - even though there is presumably no way to beat the exponential scaling.  
However, an even bigger obstacle to quantum annealing~\cite{Jorg08,Bapst13}
is posed by the fact that NP-hard problems have quantum first order
transitions~\cite{Dobrosavljevic_1990,Cugrsa1} %as typically happens for models with one-step replica symmetry breaking equilibrium solutions, 
(see Sec.~\ref{sec:1stepsystems}). 
%A rather strong obstruction  to finding the ground state of optimization problems by quantum annealing, however, is the ubiquituous first order quantum transition that arises in strong transverse fields.

%\subsubsection{Nonergodic delocalized phases in mean field quantum glasses}

A related dynamical question arises for an isolated  quantum spin glass  -- 
the setting considered in MBL: after preparation in a deep valley,
can it tunnel to others or does it remain localized? We  discuss it in Sec.~\ref{sec:qpspin}.  

\section{Replica treatment of quantum mean-field glasses}
\label{replicatreatment}

Most theoretical insights on quantum glasses have been obtained from mean-field $SU(M) $
models, where all $i=1,\dots, N$ spins interact with all others via random 
couplings with variance proportional to $J^2$ and conveniently scaled with $N$
and $M$. 
{While this looks remote from any real physical glass, {long range} couplings can be emulated 
via light-matter coupling in multi-mode cavities, see 
Sec.~\ref{sec:cavity}.  Their {\it equilibrium} properties were derived with
a replicated imaginary-time path integral formulation of the partition function~\cite{BrayMoore80}.
A saddle-point evaluation allows one to define the glass order parameter\footnote{The normalization factor $M^2$ captures the leading dependence for $M\to \infty$. In general one should naturally divide by the number of spin components ($M^2-1$).}
\begin{equation}
Q_{ab}(\tau) = 
{\frac{1}{N M^2}} 
\sum_i [\langle {\rm T} \,  {\boldsymbol \sigma}_i^a(\tau) \cdot {\boldsymbol \sigma}_i^b(0)\rangle]
\end{equation}
where $a,b=1, \dots, n$ are the replica indices, $\tau$ the imaginary time and ${\rm T}$ the time-ordering operator.
Depending on the complexity of the model, the free-energy density can 
either be just a functional of the saddle-point $Q_{ab}(\tau)$, or the partition function can be transformed 
into one for a single spin with  $Q_{ab}(\tau)$ self-consistently given by the expectation value of 
its imaginary-time correlations (see, e.g.,~\cite{Krzakala-etal}).  In all cases, $Q_{a\neq b}$ are $\tau$-independent,
vanish in the disordered phase, and adopt an RSB structure in the glass phase.
The diagonal elements $Q_{aa}(\tau) \equiv q_d(\tau)$ are independent of the replica index but 
depend on time, being $\beta\hbar$ periodic, in the whole phase diagram.
In certain cases (typically models in which also the single-spin partition function can be calculated exactly), 
$Q_{ab}$ satisfies a generic differential equation of the form 
\begin{equation}
{G_0^{-1}(\tau)}_{ac}  Q_{cb}(\tau) = \delta_{ab} \delta(\tau) + \int_0^{\beta\hbar} d\tau' \; \Sigma_{ac}(\tau-\tau') Q_{cb}(\tau')
\; ,
\label{eq:generic}
\end{equation}
with a sum over repeated indices.
${G_0^{-1}}$ is a differential operator, typically diagonal in replica indices, and its order (first or second time derivative, further time dependence) depends on the kind of degrees of freedom
and whether the system is coupled to a bath (microscopic dynamics). 
$\Sigma$ is a self-energy which typically depends on time only {\it via} $Q$. Note though that this equation is not causal, and is simpler to solve in frequency space.
We will specify $G_0^{-1}$ and $\Sigma$ in several concrete cases below.
The physical properties follow from $Q_{ab}$. Of 
special importance is the out-of-phase local susceptibility
\begin{equation}
\label{spectralfunction}
\chi_{\rm loc}''(\omega) = %\MM{\frac{1}{M^2-1??}}
{\rm Im} \, \tilde q_d(\omega + {\rm i} 0^+) 
\; , 
\end{equation}
with $\tilde q_d$ the Fourier transform of $q_d$. The right-hand-side is the 
analytic continuation to real $\omega$ of  $\tilde q_d(\omega_n)$, with $\omega_n$ the Matsubara frequencies.
%\MM{and $M$ is the number of spin components? check standard definition, cf also next Eq}. 
$\chi''_{\rm loc}$ contains essential information on the spectrum of collective excitations since
\begin{eqnarray*}
%\label{spectralfunctionLehmann}
\chi_{\rm loc}''(\omega)  = \frac{\pi}{N} 
\sum_{i m} 
| 
\langle \psi_m | 
\sigma_{i}^{1} | \psi_0 \rangle |^2 
\left[\delta(\omega - E_m+E_0) - \delta(\omega + E_m-E_0)\right], 
\end{eqnarray*}
 (at $T=0$)  {where $(\psi_m,E_m)$ are the many-body eigenstates and energies, with $0$ referring to the ground state.}
A continuous glass transition is identified by the condition $J \chi_{\rm loc} =1$,
with $\chi_{\rm loc}$  the zero frequency limit of the local susceptibility,  
$\chi_{\rm loc} = \int_0^{\beta\hbar} d\tau \, q_d(\tau)$.

\section{Classification of mean-field quantum glasses}
\label{sec:classification}

The {local degrees of freedom}, their interactions and the
 environment they couple to define different mean-field classes, 
 distinguished by their phase diagrams and properties. We list 
  three distinguishing aspects of spin models and discuss  them  in turn below.
%of different quantum glasses depend on a variety of distinguishing characteristics

\vspace{-0.2cm}

\begin{enumerate}
\item[--]
%As is well-known from the theory of classical glasses, 
The number of interacting spins decides 
on the organization of phase space as elucidated, for instance, by the replica analysis and its symmetry breaking scheme. 
%In the quantum realm 
This distinction drastically affects the nature and the order of the quantum glass transition.  

\item[--]
The symmetry group in spin space is important. It matters whether the various spin-spin interaction terms mutually commute 
(as in Ising and rotor models) or not (as in Heisenberg systems). In the latter case quantum fluctuations
are substantially stronger, entailing 
a weaker glassy order and  softer excitations, both at low and high~$T$.

\item[--]
The low-frequency dynamics depend  on whether the quantum glass is  embedded in an insulating, 
gapped host, or couples to a gapless bath with Ohmic spectrum, as in metallic glasses. 
For a given environment -- insulating or metallic --   the dynamically relevant states of all mean-field models, in spite of  quantitative differences, display a  striking universality  in their low-frequency response within the glass phase
%, the density of states of collective modes, and the associated spectral function and specific heat, 
(Sec.~\ref{sec:Landau}).
\end{enumerate}

Note that these spin glass models do not cover all possible glassy {quantum} systems. Some 
other cases are pinned elastic systems (including vortex lattices in superconductors,
{Wigner} crystals, charge and spin-density waves or disordered liquid crystals) in which {frustrating} disorder originates, {\it e.g.},  
from substrate impurities~\cite{CGL06}. 
The interested reader may consult Ref.~\cite{Giamarchi}.  Nonetheless,
once coarse-grained at the collective pinning scale, these systems resemble short-range coupled spin systems in random fields.
% BTW: the RSB structure there depends on the internal dimension $d$ (1step for d<2, I think, full RSB above). One might ask whether there is a similar reasoning as we gave for $p$-spins, which links to the bounded/unbounded Gaussian fluctuations in these dimensions  

\subsection{Pair vs multi-spin interactions}
\label{sec:pair-vs-muti}

\subsubsection{Pair interactions ($p=2$) -  continuous glass transition}

Usually one considers Ising-like interactions when discussing the distinction between pair and multi-spin 
interactions. Ising-type glasses arise in numerous contexts, as they describe generic interacting two level systems (TLS).
The  simplest representative is the transverse field Ising model (TFIM)
\begin{equation}
\hat {\mathcal H}= -  
\sum_{ij} J_{ij} \, \hat \sigma_i^z  \hat\sigma_j^z
+ \sum_i \, {\mathbf H}_i  \cdot \hat {\boldsymbol \sigma}_i 
\label{Ising-generic}
\end{equation}
where $\hat {\boldsymbol \sigma}_i = (\hat \sigma^x_i, \hat \sigma^y_i, \hat \sigma^z_i)$, 
with the usual Pauli matrices acting in the Hilbert space of the $i$'th TLS. 
In finite dimensions the first sum extends over nearest neighbours of a 
lattice.  The interaction strengths 
$J_{ij}$ are drawn  from a probability distribution, $P(J_{ij})$, 
typically chosen to be Gaussian with zero mean 
$[J_{ij}]=0$ and variance $[J_{ij}^2]=J^2/(2c)$. $J$ is  $O(1)$ and 
$c$ is the connectivity of the lattice. Mean values over $P$ are indicated with
square brackets. The ${\mathbf H}_i$ are local fields. 
%With a suitable rotation the transverse components can be reduced to $H_i^{x}$, and 
Most often one restricts to a homogeneous transverse field, ${\mathbf H}_i  = \Gamma {\mathbf e}_x $. 
Since the last term does not commute with the interactions, it induces non-trivial quantum dynamics. In the absence of longitudinal fields $H_i^{z}$, the Hamiltonian possesses an Ising symmetry %(it commutes with $\prod_i \hat \sigma_i^z$), 
which is spontaneously broken at sufficiently low $T$ and weak transverse fields $\Gamma<\Gamma_c$.
For $H_i^z\neq 0$ (as in many experimental systems of interest) the only potential phase transition  
is an RSB one, which however occurs only in high enough dimensions.
% whereby the critical dimension may depend on the range of the interactions.
%(The critical dimension for the existence of such an Almeida-Thouless transition is debated, and most likely depends on the nature of the decay of the interactions at large distances, short range, dipolar, Coulomb, etc.).

Allowing each spin to interact with all others, $c\to N-1$, places the spins on a complete
graph.  The scaling of the variance, $[J_{ij}^2]\approx
J^2/(2N)$, ensures a non-trivial thermodynamic limit, $N\to\infty$. The Hamiltonian becomes the quantum
extension~\cite{BrayMoore80} of the  
Sherrington-Kirkpatrick (SK) spin-glass (TFSK), which  is tractable with the replica trick.
{Again, only the replica diagonal, 
%The analysis is simplified by the   fact that only the diagonal terms in the replica matrix 
$q_d(\tau) \equiv Q_{aa}(\tau) =[ \langle {\rm T} \sigma^z_a(\tau) \sigma^z_a(0)\rangle ] $ depends on the imaginary time
$\tau$, while 
the RSB structure of the off-diagonal elements  
remains basically the same as in the classical limit.} Glassy freezing is signalled by a finite Edwards-Anderson parameter $q_{\rm EA}\equiv \lim_{\tau \to \infty}q_d(\tau)$ (at $T=0$). 
The resulting single-site {problem}, with the coupling to the rest encapsulated self-consistently in 
$Q_{ab}$, is of similar difficulty as the impurity problems
of Dynamical Mean Field Theory~\cite{Parcollet}. 

From the stability criterion $1= J \chi_{\rm loc}$, 
%with  the local spin susceptibility  $\chi_{\rm loc} = \int_0^{\beta\hbar} d\tau \, q_d(\tau)$, 
a $(T/J, \, \Gamma/J)$ phase diagram with a second order 
phase transition between a paramagnetic and a spin-glass 
phase is predicted~\cite{BrayMoore80}. Quantum fluctuations depress the transition temperature but do not destroy the transition. 
For $\Gamma\to 0$ the classical SK $T_c=J$ is recovered and for $T\to 0$ a quantum critical point at  $\Gamma_c=1.52J$ is found~\cite{GrempelRozenberg1998}.
Order sets in when the entropy loss and/or the loss in transverse field energy are compensated by the gain in interaction energy. 
Close to the phase transition both scale like $O(m^2)$, where $m\ll 1$ is a typical ordered moment of a single spin. 
The glass transition, whether classical or quantum, can  thus be viewed as a condensation into the first magnetization mode to become unstable.
{The glass transition differs crucially from standard ordering phenomena, however, because a large number of modes turn soft almost simultaneously. This entails many local minima with different ordering patterns. Moreover, {\em throughout} the entire glass phase there are permanently some gapless, critical modes at the verge of condensing.}
 
Interestingly enough, 
a  ``static approximation'' in which $q_d(\tau)$ is replaced by a  $\tau$-independent variational parameter captures the essence of the phase transition. 
%%\MM{cite Goldschmidt, Kopec/Usadel?}% T. K. Kopec, J. Phys. C 21, 297 (1988); K. D. Usadel ei al. , Phys. Rev. B 44, 12583 (1991) ? \cite{GoldschmidtLai1990} ?
A more detailed dynamical analysis for the  Ising case~\cite{MillerHuse1993} and related rotor models with $M\gg 1$ components~\cite{YeSachdevRead93}  shows that  the spectral gap closes as  $\Delta \sim [(\Gamma-\Gamma_c)/\ln(\Gamma-\Gamma_c)]^{1/2}$ at $\Gamma_c$, assuming a gapless spectral function $\chi''_{\rm loc}(\omega)\sim \omega$, associated with algebraic decay in imaginary time, $q_d(\tau) \sim 1/\tau^2$. 

Determining $Q_{ab}$ and $q_d(\tau)$ deeper in the glass phase requires 
the solution of  a quantum impurity problem in a frozen field
whose distribution is controlled by full RSB in the off-diagonal $Q_{ab}$~\cite{AndreanovMuller}. The latter ensures marginal stability, which in turn implies a gapless spectrum  {\em everywhere} in the glass phase.  
Moreover, the limit of small $\Gamma$ admits a scaling form for
the mean field solution. From this it follows that the low-frequency spectral function becomes 
independent of $\Gamma$,  
\bea
\label{spectralfunctionTFSK}
\chi_{\rm loc}''(\omega; \Gamma\ll J) = 0.59 \,\omega/J^2, 
\eea
which we will interpret physically in Sec.~\ref{sec:Landau}.

\subsubsection{$p>2$ - discontinuous onset of glassy order}
\label{sec:1stepsystems}

Models in which   $p>2$ classical spins interact simultaneously are particularly interesting since at the mean field level their dynamics are described by equations identical to those arising in the Mode-Coupling Theory of structural glasses. 
It is therefore believed that they capture the physics of the structural glass transition and the glassy phase~\cite{Bouchaud,BerthierBiroli}.  
Moreover, such multi-spin Hamiltonians are ubiquitous in combinatorial optimization problems when translated into questions about ground states of statistical mechanics systems. Defined on diluted graphs, the $p$-spin models are, 
{\it  e.g.}, closely   connected to the $K$-satisfiability
 problem~\cite{Monasson}. 
  
A typical multi-spin Hamiltonian reads~\cite{Goldschmidt,Niri98,Cugrsa1,Cugrsa2,bath-spherical,bath-SU2} 
\begin{equation}
\hat {\mathcal H} = - \sum_{i_1 \neq \dots \neq i_p} J_{i_1\dots i_p} 
\hat \sigma^z_{i_1} \dots \, \hat \sigma^z_{i_p} 
+ \sum_i \Gamma_i \hat \sigma_i^x 
%+ \sum_i h_i \hat \sigma_i^z
\; , 
\end{equation} 
with the  parameter $p$ taking  any integer value $p \geq
3$. The model is defined on a hypergraph, the sum running over all $p$-uplets.  
The couplings are 
random independent variables with variance $p!{J}^2/(2 N^{p-1})$.

The important difference w.r.t. pairwise interacting models lies 
in the scaling of the loss of entropy or transverse field energy as compared to the interaction energy gain with a putative small emerging 
magnetization pattern $m_i= \langle \sigma_i^z\rangle$. 
The former still scale as $O(m^2)$, but the energy gain is only $O(m^p)$ and cannot compensate for this loss if $m$ is small. 
The only possibility for a phase transition is that the order parameter, $q_{\rm EA}= 1/N \sum m_i^2 $ jumps discontinuously to a finite value at a critical point. 
%In the classical limit, 
Technically, this is reflected by a one-step RSB.  
Part of the liquid entropy is transferred to configurational entropy without a thermodynamic phase transition (the free energy is smooth). 
The Gibbs weight distributes over an exponential number of metastable states.
When quantum fluctuations are switched on, and temperature is sufficiently low, the transition necessarily has to change 
nature since the configurational entropy cannot contribute to the free energy. 
The phase transition thus becomes truly first order~\cite{Bicu} and the ground states on the two sides of the transition are essentially unrelated. For this reason quantum annealing is not expected to be useful for $(p>2)$-spin systems and other systems with a 1-step RSB. 

In quantum and classical glasses of this type alike, the proliferation of metastable states affects the dynamics, but does not necessarily change the static observables. 
{Over a finite range of free-energy densities one finds exponentially many local minima, the fewer the lower the free-energy.} 
As temperature decreases the Gibbs weight concentrates on lower energy states.
A thermodynamic freeze-out transition occurs once the lowest $O(1)$ states dominate. Consequently, }the dynamical glassy phase extends beyond the  thermodynamic one~\cite{Cuku93,Culo,Culo1}. 

The $p$-spin potential was also used to  model a quantum 
particle in a random potential, in which the interactions are between $p$-uplets of coordinate positions
on an $N$-dimensional sphere. Conventional kinetic energy is given to the particle. In these cases an equation of the kind of Eq.~(\ref{eq:generic})  
applies with 
${G_0^{-1}}_{\!\! ab} = (d_\tau^2 + \mu) \delta_{ab}$ with $\mu$ a Lagrange multiplier enforcing the spherical 
constraint and $\Sigma_{ab} = J^{2} Q^{\bullet (p-1)}_{ab}$ where $\bullet$ denotes a normal power. Self-energies
of this very same kind appear in the SYK model, Sec.~\ref{sec:SYK}.

\subsection{Commuting vs non-commuting interactions}

\subsubsection{Commuting interactions}

The disordered quantum rotor model  is defined by~\cite{YeSachdevRead93}
\begin{equation}
\hat {\mathcal H} =
\frac{g}{2M}
\sum_i \hat {\mathbb L}_i^2  
+\frac{M}{\sqrt{N}} \sum_{i<j} J_{ij} \, \hat {\mathbf n}_i \cdot \hat {\mathbf n}_j 
\qquad
\;\;
\hat {\mathbf n}_i^2 = 1 \;\; \forall i 
\; . 
\end{equation}
The $M$ components $\hat n_i^\mu$ of the $i$th unit-length rotor $\hat {\mathbf n}_i$ 
 commute with each other, unlike the components of quantum spins. As a consequence, 
all interaction terms mutually commute.
$\hat L_i^{\mu\nu}$ (with $\mu<\nu$, $\mu, \nu =  1, \dots, M$) are the $M(M-1)/2$ components of the 
angular-momentum generator $\hat {\mathbb L}_i$ in rotor space, and
$[L_i^{\mu\nu},n_j^\sigma]=i \delta_{ij} (\delta_{\mu\sigma} n_j^\nu - \delta_{\nu\sigma} n_j^\mu )$.
 The $J_{ij}$ are $O(1)$ randomly distributed uncorrelated exchange
constants. 
 %\MM{NB: note that there is no single $L$ operator for $M=1$!}
As $g\to 0$ the model reduces to the classical, infinite-range, $M$-component 
spin glass, the limit $M\to 1$  being very similar to the Ising model.
 
There is no Berry phase in the real action of the path integral representation of the
partition function, and the $O(M)$ symmetric saddle-point~\cite{YeSachdevRead93} in the $M\to\infty$ limit 
yields a self-consistency equation, Eq.~(\ref{eq:generic}),
with ${G_0^{-1}}_{\!\! ab}  = g^{-1} (d^2/d\tau^2 + \mu) \delta_{ab}$, where $\mu$ enforces the constraint $\hat {\mathbf n}_i^2 = 1$,
and one has $Q_{ab} = q_d \delta_{ab}$ and $\Sigma_{ab} = Q_{ab}$.
%This equation is of the generic kind (\ref{eq:generic}), for 
In the glass phase the same equation is obeyed by $q_d(\tau)-q_{\rm EA}$, while $Q_{a\neq b}=q_{\rm EA}$ becomes non-zero, without however breaking the replica symmetry.   
The integro-differential equation is easy to solve in frequency space and yields spectral functions that we will review in Sec.~\ref{sec:Landau}. The quantum critical behavior was found to be in the  
TFSK universality class, which was rationalized by showing that $1/M$ corrections  do not modify the critical exponents and the low frequency spectrum.

%\cite{GoldschmidtLai1990}? Gamma_c from static approx
% Independence of results on large $M$
%$\chi''_{\rm loc}(\omega)\sim \omega$ in the glass phase, 
%\textcolor{red}{end to be edited}

\subsubsection{Non-commuting interactions - $SU(M)$ spins}
\label{SUM}

The most relevant model in this class is the quantum Heisenberg spin glass,
\begin{equation}
\hat {\mathcal H}= -  
\sum_{i\neq j} J_{ij} \, \hat {\boldsymbol \sigma}_i \cdot \hat {\boldsymbol \sigma}_j
\; . 
\label{Heisenberg}
\end{equation}
A static approximation used to evaluate the instability condition for $S=1/2$~\cite{BrayMoore80}
yields a surprisingly good estimate for $T_g\approx \sqrt{3}/12 \approx 0.14 \, J$, obtained with Quantum Monte Carlo calculations~\cite{GrempelRozenberg1998_2}. The substantial reduction with respect to the classical $T_g =  J/4$ is due to quantum fluctuations. 

With the static approximation, though, one cannot access the spin dynamics. A crucial step forward was taken in~\cite{SachdevYe93} where $SU(2)$ was promoted to $SU(M)$
and a Schwinger boson representation of the spins, with the number of bosons constrained to 
be $n_b=SM$, was adopted.
A first analysis in the 
$(M,S)$ parameter space was performed in the large $M$ limit. 
Soon after, a complete solution 
with $M\to\infty$ and fixed $S$
was derived, and global aspects, some of them also valid for $SU(2)$, discussed~\cite{ParcolletGeorgesSachdevPRL,ParcolletGeorgesSachdevPRB}. The 
solution is formulated in terms of the Green functions of the bosons, $G^{ab}_B(\tau) 
\equiv - M^{-1} \sum_\mu [\langle {\rm T} \hat b_\mu^a(\tau)\hat b_\mu^{b\dag}(0) \rangle]$ 
and their self-consistency equations
are again of the form of Eq.~(\ref{eq:generic})
with 
\begin{equation}
\begin{split}
%G^{ab}_B({\rm i} \omega_n) &= [-{\rm i} \omega_n +\lambda - \Sigma^{ab}_B({\rm i} \omega_n)]^{-1}\; , \\
%%\frac{d G^{ab}_B(\tau)}{d\tau} &= 
%%- \mu^a \delta^{ab} G^{ab}_B(\tau) + \int_0^{\beta\hbar} \!\!
%%d \tau' \, \sum_c \Sigma_B^{ac}(\tau-\tau') G_B^{cb}(\tau') + \delta(\tau)
%%\; , 
%%\label{eq:dG/dt}
%%\\
{G_0^{-1}}_{\!\! ab} = (d_\tau + \mu^a) \delta_{ab}
\; , 
\qquad
\Sigma_B^{ab}(\tau) &= J^2 G^{ab}_B(\tau)G^{ab}_B(\tau)G^{ab}_B(-\tau)
\; . 
\end{split}
\end{equation}
$\mu^a$ is a chemical potential which fixes $G_B^{aa}(\tau=0)= -S$.
The disorder-averaged spin-spin correlator is obtained from $G_B$ by 
\bea
\label{Q=GG}
{q_d}(\tau) 
%[=\chi_{\rm loc}(\tau) ]
\equiv 
\frac{1}{M^2}
[\langle \hat {\boldsymbol \sigma}^a(\tau) \cdot 
 \hat {\boldsymbol \sigma}^a(0)\rangle] = G_B^{aa}(\tau) G_B^{aa}(-\tau)
\; .
 \eea 
For large $S$ one finds an essentially classical glass transition at $T_g \sim S^2 J$, where the bosonic spinons condense. The 4-spinon 
interaction term resembles the $p=4$-spin problem, and for similar reasons one finds a 1-step RSB transition.
 %symmetry breaking transition is 1-step (with $T_d\approx 2/3\sqrt{3} JS^2$ - which shows that $S\to 0$ would correspond to a quantum critical point). 
%arises 
This leaves open the choice of the size of replica blocks, which selects the energy of the targeted metastable states. Usually one uses  the condition of marginal stability (vanishing of the ``replicon'' mode), as those are reached dynamically~\cite{BiroliParcollet02} and exhibit a gapless spectrum.
Extremization of the free-energy 
density instead describes dynamically inaccessible states with a gap in
$\chi_{\rm loc}''(\omega)$. 

At large $S$, quantum fluctuations introduce a frequency scale $\sim S J$, which sets the crossover temperature where collective quantum dynamics emerges. Consistent with the Landau analysis of Sec.~\ref{sec:Landau} marginal states are found to have $\chi''_{\rm loc}(\omega)\sim \omega$ and a low-$T$  specific heat that scales as $T^3$~\cite{Schehr04,Schehr05}
(contrary to the linear $T$ dependence originally claimed in~\cite{ParcolletGeorgesSachdevPRL,ParcolletGeorgesSachdevPRB}).
% We will come back to this issue when discussing the SYK phenomenology in Sec.~\ref{sec:SYK}. 

The case of small $S\ll 1$ exhibits much  stronger quantum fluctuations and is thus more interesting. 
Here, the  spin-fluid, paramagnetic regime survives down to much lower temperatures $ T\ll  J$ and is 
  radically different from the classical paramagnet  of the high $T$ and large $S$ limits. 
%At high $T$, the spins behave as local moments and $Q(\tau) \simeq S(S+1)$. 
%Lowering $S$ and $T$ below $JS^2$, the Curie temperature is reduced to $S^2$ by quantum fluctuations. 
Below $T\sim J$,  the system enters a gapless, quantum critical regime in which
the Green function assumes long-time tails, $G^{aa}_B(\tau)\sim 1/(J\tau)^{1/2}$, while the melon-bubble self-energy is unusually large, $\Sigma(\omega) \sim (\omega J)^{1/2}\gg \omega$, reflecting the  strong  scattering among spinons. 
%Essentially the same is found with fermionic representations which will be further discussed in Sec.~\ref{sec:SYK}.
From  the spin-spin correlation $q_d(\tau)\sim 1/(J\tau)$ one extracts a  susceptibility very similar to the one of marginal Fermi liquids:
%Analytic continuation to real frequencies yields a very similar 
 $J \chi''_{\rm loc}(\omega) \simeq \tanh(\hbar\omega/2T)$
implying $\chi' _{\rm loc}\simeq \ln(J/T)$. One further finds linear in $T$ specific heat and a residual low-temperature entropy. This holds in the paramagnetic regime
 $T_g  < T < J$, for an extended range of $S$ all the way up to 
$S \simeq 1$. These properties are very similar to those of black holes to which this model and  SYK is related.
%welcome properties  for the gravity applications of SYK.

From Eq.~(\ref{Q=GG}) one sees that the onset of glassy order  modifies the Green functions beyond the time scale $\tau^* = (\omega^*)^{-1} = (q_{\rm EA}J)^{-1}$ where $G_B(\tau)G_B(-\tau) \sim q_{\rm EA}$. At that scale the spinons get confined, and  $Q(\tau)$ crosses over to $Q(\tau) -q_{\rm EA} \sim \tau^*/(J\tau^2)$.
This corresponds to a linear spectral function $J\chi_{\rm loc}''(\omega)\sim \omega/(q_{\rm EA}J)$, as is found in all insulating mean-field glasses, cf. Sec.~\ref{sec:Landau}. Note, however, that weak glass order implies a potentially large prefactor $\sim 1/q_{\rm EA}$, reflecting very slow collective modes.

The bosonic representation is {\it a priori} best suited for large $S$. Fermionic $SU(M)$ representations 
are instead believed to capture better the physically most relevant case $S=1/2$.  
Such an approach yields essentially identical results for the paramagnetic phase~\cite{ParcolletGeorges}, though no glass transition is found at $M=\infty$ in this case~\cite{SachdevYe93}, since quantum fluctuations are much stronger. Yet, a $1/M$ expansion~\cite{ChristosSachdev22} yields an instability to a glass phase where a fermionic bilinear condenses at $T_g\sim J \exp(-\sqrt{\pi M})$ (now with continuous RSB -- as the effective action is quadratic in the emerging order parameter, like in $p=2$ models). One expects a similar suppression of the order parameter, $q_{\rm EA} \sim T_g/J$ (which ensures that at $T \sim T_g$ the low-$T$ linear spectral function matches that of the paramagnetic regime).  %J \chi''_{\rm loc}(\omega) \simeq \tanh(\hbar\omega/2T)$
Exact diagonalization for all-to-all coupled $SU(2)$ spins~\cite{ArracheaRozenberg01, Shackleton2021} indeed yields $q_{\rm EA}\approx 0.02$. 
%This implies a large prefactor in the linear low frequency spectral weight and very slow collective modes. 
However, currently accessible system sizes of $\sim$ 20 spins do not allow to disentangle the regime $\omega<\omega^*$  from the broadened peak $\sim q_{\rm EA}\delta(\omega)$  reflecting static spin glass order. Nonetheless, they do exhibit spectral features as predicted by the fractionalized large $M$ approach for $\omega>\omega^*$.

\subsubsection{Heisenberg glasses and Many-Body Localization}

We close this section by remarking 
that Heisenberg spin chains with coupling $J$ under strong, but random {\it i.i.d.} 
local  fields ${\mathbf H}_i = H^x_i {\mathbf e}_x$ drawn from a distribution of width $W \gg J$, 
has become one of the standard models for MBL~\cite{Nandkinshore}. Indeed, the typically large mismatch between the local gaps $H_i^x$ cannot be  bridged resonantly by the weak couplings $J$, and thus coherent, bath-free dynamics is expected to remain stuck close to an arbitrary initial state. 
It is worth pointing out again that MBL is favored by completely opposite ingredients than a spin glass phase: it requires the non-commuting interactions to be weak, and the dimensionality should be low. In contrast, in the parameter range where a glass with non-trivial energy landscape exists, one expects rapid and full thermalization within any local minimum of the energy landscape, independently of the coupling to a bath.

\subsection{Insulating vs metallic environment}

Remarkably,  all insulating mean-field quantum glasses in their marginally stable glassy states exhibit the same linear scaling $\chi''_{\rm loc}(\omega)\sim \omega$, albeit with quantitatively different prefactors,  and a  low-$T$ specific heat $c_V\sim T^3$. This contrasts with a spin glass coupled to a gapless bath, {\it e.g.} 
via Kondo coupling to itinerant conduction band electrons, as in heavy fermion materials~\cite{SenguptaGeorges}. The dissipation 
due to the low-energy bath degrees of freedom  slows down the collective modes and thereby substantially enhances the low-frequency spin spectral function and the spin contribution to the specific heat, as we will review via a Landau approach in the next section.

\section{Landau theory \& the low frequency spectrum}
\label{sec:Landau}
%\subsubsection{Spectral function close to the transition}

A Landau expansion was reviewed in \cite{SachdevRead96} and we only sketch it here.
The aim is to construct an effective action for the glass order parameter 
$Q_{ab}$ by integrating out all other degrees of freedom. 
This approach 
%in the Ginzburg-Landau-Wilson spirit of an order parameter built from the elementary spin degrees of freedom 
is expected to succeed for phase transitions in which the spins undergo standard ordering, as {\it e.g.} in the TFIM and rotor models. However, in Heisenberg-type models, where at least for $M\gg 1$  the spins fractionalize into deconfined spinons in the paramagnetic phase, it will likely  fail to capture the entire dynamical crossover functions correctly. Methods as the ones for deconfined quantum critical points~\cite{Senthil} may instead be required. 
Still, even in these cases the Landau approach may capture well 
the glassy phase in its low-frequency, spinon-confined regime.

Considering symmetry restrictions and constraints on the replica structure, and assuming the absence of time-reversal symmetry-breaking fields, the resulting mean-field action close to the quantum glass transition takes the form~\cite{ReadSachdevYe95}
\bea
\label{action}
&& 
n\beta {\cal F}[\delta Q]= 
\sum_a \int d\tau \left. {\cal L}[\delta Q_{aa}] \right.
%(\tau_1,\tau_2)]\right|_{\tau_1=\tau_2=\tau}
%\int d\tau \left.(r+P(\p_{\tau_{1,2}}))\delta Q_{aa}(\tau_1,\tau_2)\right|_{\tau_1=\tau_2=\tau}
%\sum_{\omega_n} (r+M(\omega))\delta Q_{aa}(\omega_n)+
%u\sum_a\left[\sum_{\omega_n} Q_{aa}(\omega_n)\right]^2
+u\sum_a\int d\tau  \left[{\delta} Q_{aa}(\tau,\tau)\right]^2
\nn\\
&&
\;\;\;
-c_3 \sum_{abc}\int d\tau_1 d\tau_2 d\tau_3 \delta Q_{ab}(\tau_1,\tau_2)\delta Q_{bc}(\tau_2,\tau_3)\delta Q_{ca}(\tau_3,\tau_1) +O(\delta Q^4)
\; ,\nn
\eea
where
$\delta Q(\tau, \tau')=Q_{ab}(\tau, \tau')-c\delta_{ab}\delta(\tau-\tau')$, with the constant $c$ fixed to remove the uninteresting and non-universal short-time dynamics.
The coefficients are bare correlations of the degrees of freedom coupling to $Q$, but only in the linear term their time-dependence is relevant. The leading low-frequency dependencies, 
distinguish whether the degrees of freedom are gapped (insulating) or gapless like in a Fermi sea (metallic), 
and are 
\begin{equation}
\begin{aligned}
{\cal L}_{\rm ins}[\delta Q_{aa}] &=
(r+ \partial_{\tau_1}\partial_{\tau_2}) \delta Q_{aa}(\tau_1,\tau_2) 
\; , 
\\
{\cal L}_{\rm met}[\delta Q_{aa}] &= \left[r' - (\tau_1-\tau_2)^{-2} \right] \delta Q_{aa}(\tau_1,\tau_2)
\; ,
\end{aligned}
\end{equation}
where ${\tau_1=\tau_2=\tau}$.
%where the long time tail in the metallic case reflects the gapless, Ohmic particle-hole excitations of the metallic bath. 
In the metallic case we dropped the subleading time derivatives, obtaining a leading $|\omega_n|$ dependence in Matsubara space, while
the coefficient in insulating glasses displays a faster $\omega_n^2$ dependence.

\subsection{Insulating glasses}

Minimizing ${\cal L}_{\rm ins}$ with respect to $\delta Q$ close to the quantum transition at $T=0$  yields~\cite{ReadSachdevYe95}  
%\bea
%\label{insulatingSpectrum1}
$\delta \tilde Q_{aa}(\omega_n)=-\sqrt{\omega_n^2+\Delta^2}$
%\eea
(dropping numerical coefficients), which implies the spectral function 
\bea
\chi''_{\rm loc}(\omega)=\textrm{sgn}(\omega)\sqrt{\omega^2-\Delta} \, \Theta(|\omega|-\sqrt{\Delta})
\; .
\eea
The parameter-dependent energy $\Delta$ plays the role of a spectral gap,  which closes, $\Delta \rightarrow 0$, upon approaching the  transition from the paramagnetic side.
However, remarkably, within the glassy phase the gap remains pinned to zero, $\Delta=0$, which entails a linear low frequency spectral function:
\bea
\label{insulatingSpectrum}
\chi_{\rm loc}''(\omega)=\omega
\; .  
\eea
%which is indeed found in a large variety of solvable insulating glasses.
This dynamical crossover agrees with the exact solution of large $M$ quantum rotors and the critical behavior of the TFIM. 
Allowing for RSB, one finds marginally stable solutions with a vanishing replicon and gapless spectrum. %This also holds even deep in the glass phase of the Ising model.  

Note that, as anticipated above, the high-frequency spectrum predicted 
by the Landau approach differs qualitatively from the spin susceptibility 
$\chi''_{\rm loc}(\omega)\sim {\rm sgn}(\omega)$ of the large $M$ limit of Heisenberg models, 
which hinges on spinon deconfinement. 

\subsection{Metallic glasses}

Minimizing ${\cal L}_{\rm met}$ close to the transition one finds~\cite{SachdevReadOppermann95,SenguptaGeorges},
%\bea
%\label{metallicSpectrum1}
$\delta Q_{aa}(\omega_n)=-\sqrt{|\omega_n|+\Upsilon}$,
%\; ,
%\eea
whereby  the crossover energy $\Upsilon$ vanishes  again  at the quantum 
transition, $\Upsilon\rightarrow 0$, and stays pinned to $\Upsilon=0$ in the glassy phase.
Here the spectral function is everywhere \textit{gapless} and approaches
\bea
\label{metallicSpectrum}
\chi''_{\rm loc}(\omega)=\frac{1}{\sqrt{2}}\frac{\omega}{\sqrt{\Upsilon+\sqrt{\omega^2+\Upsilon^2}}}
\overset{\Upsilon\to 0}{\longrightarrow}
{\rm sgn}(\omega)\frac{|\omega|^{1/2}}{\sqrt{2}}
\eea
in the glass phase,
which, at low frequencies,  is substantially stronger than the one of insulating glasses, Eq.~(\ref{insulatingSpectrum}). 
%\MM{may be dropped: It was noted that the non-linear susceptibility has an anomalously weak singularity at the transition from metallic paramagnet to metallic spin glass.} 
Again, the Landau approach does not capture the high-frequency features associated with spinon deconfinement~\cite{Shackleton2021}. 

\subsection{Physical interpretation of glassy spectral functions}

The low-frequency spectral functions in the glass phase can be interpreted as arising from of set of collective random spin-density modes that behave as either underdamped or overdamped harmonic oscillators.

In the glass phase a local minimum of the free-energy landscape (expressed as a functional of local magnetizations) has a Hessian with positive eigenvalues 
$\lambda_k$ which may be interpreted as generalized spring constants. The marginal stability of the relevant glass states assures that their distribution is gapless, and random matrix theory suggests that at small $\lambda$, it behaves as $\rho(\lambda) \equiv N^{-1}\sum_k \delta(\lambda-\lambda_k) = c\sqrt{\lambda}$. 

In the insulating case one expects the Hessian normal modes to behave as independent harmonic oscillators  with an effective mass $M$, set by the typical microscopic time scale of the spin dynamics. The associated oscillator frequencies then scale as  $\omega_k =  \sqrt{\lambda_k/M}$, with a spectral density
$
\rho(\omega) \sim  cM^{3/2} \omega^2
$.
This correctly predicts the specific heat to scale as $c_V(T)\sim T^3$ at low $T$, rationalizing the result of~\cite{Schehr04} (with the notable exception of quantum particles at jamming where $\rho(\lambda) \sim \lambda^{-1/2}$~\cite{Franz2019}).
The typical mean-square displacement of oscillator mode $k$ scales as $\<x_k^2\>\sim \hbar/M\omega_k \sim \hbar/\sqrt{M \lambda_k}$, and from the Lehmann representation, Eq.~(\ref{spectralfunction}), one expects the spectral function to scale as 
\bea
\chi''_{\rm loc}(\omega)&\sim& \rho(\omega) \<x^2\>_\omega \sim  c\sqrt{M} \omega
\; ,
\eea
which rationalizes the linear scaling in Eq.~(\ref{insulatingSpectrum}). 

The random oscillator picture is further supported by the scaling of the ratio $\chi''_{\rm loc}(\omega)/\omega$.
Exact calculations for rotors~\cite{ReadSachdevYe95} or a particle in a random environment~\cite{Bicu} indeed obtained 
a ratio $\sim \sqrt{M}$. In the TFSK,  an effective potential construction yields $c \sim \sqrt{\Gamma}/J^2$ for $\Gamma\ll J$, while one expects $M\sim 1/\Gamma$. This correctly  predicts  the $\Gamma$-independent ratio  in Eq.~(\ref{spectralfunctionTFSK}).

For metallic environments, the oscillator dynamics is instead expected to be overdamped with a friction coefficient $\eta$ proportional to the prefactor of $|\omega_n| \tilde Q_{aa}$. In this case the  characteristic frequency of mode $k$ is given by the relaxation rate, $\omega_k \sim \lambda_k/\eta$, implying a 
mode density $\rho(\omega)\sim c \eta^{3/2}\omega^{1/2}$ with larger weight at low frequencies. Accordingly, one expects corrections $\delta c_V(T) \sim T^{3/2}$ to the specific heat, as indeed predicted in~\cite{SenguptaGeorges}.  
The typical oscillator displacement follows from the scaling $\hbar \omega_k \sim \lambda_k x_k^2$, that is,
$\langle x_k^2\rangle \sim \hbar/\eta$, independent of the mode.
%With the interpretation of independent damped oscillators in the glass phase one thus expects 
The spectral function should then scale as 
\bea
\chi''_{\rm loc}(\omega)&\sim&  \<x^2\>_\omega 
\rho(\omega) \sim  c \sqrt{\eta \omega}
\; ,
\eea
which again correctly reproduces the result of the Landau theory given in Eq.~(\ref{metallicSpectrum}), as well as the scaling with $\eta$. 

Unfortunately, little is known about the collective modes in short-range glasses. Those may play an important role in electronic processes such as  hopping transport in insulators, as potential glue for strong-coupling superconductors, or as inelastic scatterers affecting the electrical resistance in metallic glasses.

\section{Tunneling and eigenstate localization}
\label{sec:qpspin}

It is interesting to ask whether and how quantum tunneling allows one to explore the clusterized  landscape of quantum  models with a 1-step RSB pattern.
Having initialized a quantum Ising glass in a deep energy valley, tunnelling might not take place, 
even in the thermodynamic limit, since the remaining discreteness of the eigenstates (within one valley) may prevent resonant intervalley coupling. 

This problem was first analyzed for $p>2$ models~\cite{Baldwin17}.  Computing the  tunneling amplitude between valleys of equal internal energy $E$ and equating it to the intra-valley level spacing determines a minimal transverse field $\Gamma_c(E)$, below which wavefunctions and thus quantum dynamics, remain valley-localized.
This situation is globally non-ergodic, yet delocalized within one valley - an unusual situation, believed to be impossible in finite dimensional models of MBL.
%Deep enough in the energy landscape a threshold transverse field is required for a mean field glass to tunnel out of its initial valley. 
The 
%Scardicchio Laumann:
delocalization threshold  $\Gamma_c(E)$ was found to lie consistently below the threshold $\Gamma_d(E)$ where glassy dynamics sets in. Indeed, $\Gamma_c$ only senses the energy barriers in phase space,  while dynamics can be exponentially trapped by entropic ({\em free} energy) barriers, in spite of percolation in the energy landscape.

At finite $p$, once quantum tunneling resonantly connects different valleys of energy $E$, the eigenstates hybridize all classical configurations having energy close to $E$, since the wavefunctions behave ergodically within a valley. 
This is not so in the strict limit $p\to \infty$~\cite{Baldwin16,FaoroFI2019,BiroliTarziaVivo2021}  where the random energy model results, and each valley contains only one single configuration.
In this limit,  only states in a very narrow energy shell hybridize, its  width being the exponentially small tunneling between those states. 

It was claimed that tunneling under barriers can be exploited for a quantum algorithm to efficiently find other valleys of equal internal energy~\cite{Baldwin18}.

\section{Interplay of glassy  order with localization phenomena}
\label{sec:interplay}

%\subsection{Glassy density order and quantum localization}
%\label{sec:glassloc} 

The interplay of glassy freezing with other quantum phenomena, especially with the localization of fermions or bosons, is a rich subject. Below, we discuss how (fermionic) electron glasses become metals, and how glasses of bosons may become superfluid.
%  the Coulomb interacting metal-insulator transition, or cold atoms with cavity mediated (long range) interactions 

 %(or the quantum dynamics of quantum particles \cite{Franz2019, Artiaco21} 

\subsection{Quantum electron glasses}
\label{sec:QEG}

The classical Coulomb glass is a disordered insulator with unscreened Coulomb interactions between 
localized electrons, realized in {\it e.g.} 
doped semiconductors~\cite{EfrosShklovskii,Vaknin00,Vaknin02}. A standard model is 
\begin{equation}
{\mathcal H}_{\rm Cb} = \sum_i \epsilon_i n_i + \sum_{i<j} \frac{n_i n_j}{r_{ij}}
\label{eq:Heg}
\end{equation}
with random energy $\epsilon_i$ on site $i$ drawn from a box
distribution $[-W/2, W/2]$. The local occupation
number is $n_i=0,1$ and $r_{ij}$ is the distance
between sites $i$ and $j$. The number of electrons is a
fixed fraction of the sites, which are arranged on a lattice or at random. 

The unscreened, long range $1/r$ interactions between localized electrons enforce the Efros-Shklovskii (ES) Coulomb gap in the single-particle density of states, $\rho(E)$. 
It has long been conjectured that the stability constraint $\rho(E)\leq {\rm const.}\times E^{d-1}/e^{2d}$ (which assures that particle-hole recombinations do not lower the energy) is only marginally satisfied. A mean-field treatment 
with marginal  full RSB predicts this critical property~\cite{MuellerIoffe04,PankovDobrosavljevic05,MuellerPankov07}.  
 This phenomenon is directly analogous to the appearance of a linear density of local fields in the classical SK model, which is known to saturate similar stability bounds \cite{MuellerWyart2015}.

Upon increasing the density of dopants the hopping $t_{ij}$ between the sites $i$ becomes important and a term $\sum_{i<j} (t_{ij} c_j^\dagger c_i +h.c.)$ must be added to ${\mathcal H}_{\rm Cb} $ in 
Eq.~(\ref{eq:Heg}). 
Single-particle eigenfunctions $\psi_\alpha$ now spread over several sites. When written in the basis of $\psi_\alpha$, the interactions take the generic form $\sum_{\alpha,\beta,\gamma,\delta} U_{\alpha\beta\gamma\delta} c^\dagger_\alpha c^\dagger_\beta c_\gamma c_\delta$, which couples all wavefunctions $\psi_\alpha$ that overlap in real space. Close to the metal-insulator transition this term resembles the fermionic models reviewed in Sec.~\ref{sec:SYK}.
%Some well connected pairs of sites may host two electronic states, that can be considered as two level systems which interact via dipolar interactions. Such dipoles can in principle undergo a glass transition~\cite{dipolar-glasses}, but whether this happens before the electrons delocalize, and whether the interactions are sufficient to overcome the random fields acting on the dipoles, has not been studied carefully. 

Eventually localization breaks down and the wavefunctions become extended~\cite{Epperlein1997}. At the same time screening sets in, the effective Coulomb interactions
become short range and the density of states at the Fermi level becomes finite.
Close to criticality the growing Coulomb gap suppresses the occurrence of
strong resonances for low-energy electrons, so that their wavefunctions
concentrate less on rare paths. This results  in a larger fractal dimension as compared to the ones 
of non-interacting critical wavefunctions~\cite{Amini_2014, BurmistrovMirlin2013}. 
%However, while delocalization becomes very inefficient at the single particle level, with high resistivities at low $T$, inelastic  effects of the power law interactions do not allow for genuine  many-body localization, as was confirmed for a 1d toy model~\cite{Rademaker2020}.
%\MM{~\cite{Rademaker2020} : interplay of pseudogaps in a 1d power law interacting spin glass and MBL:  The pseudogap makes delocalization less efficient but cannot halt it of course, the glass order persists.}

Already in the proximity of the insulating state one finds multiple solutions to Hartree-Fock equations. This may signal the emergence of metastability~\cite{PastorDobrosavljevic99,Dobrosavljevic03}, which then extends all the way into the insulating electron glass. Slow relaxation and aging in gated 2d electron gases in silicon~\cite{BogdanovichPopovic2002,JaroszynskiPopovic2007} at metallic densities was taken as an indication for electron glassy behavior.
However, the existence of a genuine glass transition in $d=3$  remains debated even in the classical limit, as the random potential breaks a potential Ising symmetry so that only the replica symmetry is left to be broken. Numerical simulations have not fully clarified the situation~\cite{Palassini,AndersenKatzgraber}.

In Sec.~\ref{cavitymetallicglass}  we will discuss the related case of fermionic atoms with photon-mediated long-range interactions. Interestingly, the latter allow for a genuine glass transition (with RSB) already in the metallic phase.

\subsection{Superfluidity and glassy density order}

Experiments that had suggested superflow in solid, but possibly amorphous Helium~\cite{KimChan04} had raised the interesting question whether and how (glassy) density order of bosons could coexist with the a priori competing superfluid order. 
%Competition between density order and superfluid condensation. Originally motivated by non-classical rotational inertia observed in solid helium, which suggested a putative coexistence of a (glassy/amorphous) solid structure with a superfluid component. (Those experiments  were eventually explained by some artefacts.)
For models with mean field-like, frustrated density interactions (such as e.g. realized in multimode cavities) and locally hopping bosons, it was indeed found that  the two orders can coexist~\cite{Carleo2009}, while they try to avoid each other locally: if the local superfluid order parameter is high, the glassy density order is weak, and vice versa~\cite{YuMueller2012}. This phase is the bosonic 
analogue of the metallic glass of fermionic atoms discussed in Sec.~\ref{cavitymetallicglass}.
 For a specific 3d model whose quantum Hamiltonian maps onto the Fokker-Planck equation of Brownian hard spheres, the existence of such a superglass phase could be inferred explicitly, drawing on the knowledge on the classical glass phase of hard spheres~\cite{BiroliChamonZamponi2008}.   

\section{Experimental realizations of quantum glasses}
\label{sec:experimentalsystem}
Let us now review a few selected examples of quantum glasses in correlated materials and possible realizations in artificial structures involving light-matter coupling.

\subsection{Quantum Ising spin glasses}
\label{sec:QIsingspinglass}

For long, LiY$_{1-x}$Ho$_x$F$_4$
%, a dipolar interacting Ising system, 
was considered to be ``the'' quantum spin-glass realization~\cite{RosenbaumAeppli}. The rare earth magnetic ion Holmium (Ho) has an Ising doublet with a 
large magnetic moment of $5.4 \mu_B$ as its crystal field ground state.
As the content of  Ho is decreased below $x\approx 44\%$, the material turns from a dipolar ferromagnet to a dipolar-interacting Ising spin glass. Quantum fluctuations are introduced by a transverse field $H_t$, which at second order in perturbation theory hybridizes the two states of the Ho ground-doublet, resulting in an effective transverse field for the Ising doublet  $\Gamma \sim H_t^2/\Delta$, where $\Delta$ is the  gap in the crystal field spectrum.

Initially, the variation of the susceptibility with $\Gamma$ for $x=0.167$ was interpreted in terms of a first order transition at low temperatures~\cite{RosenbaumAeppli1}. With decreasing $\Gamma$ the linear susceptibility reflects a rather sudden slowdown of dynamics and the onset of low frequency $1/f$ noise. However, the non-linear susceptibility does not diverge, ruling out a continuous glass transition. 
A first order transition, instead, is at odds with the rule of thumb that two-body interactions usually entail continuous  transitions. It was later argued~\cite{LaflorencieSchechter} that no equilibrium glass transition can occur at large $H_t$, since the latter induces finite transverse moments whose internal longitudinal fields break the Ising symmetry explicitly. In the absence of a de Almeida-Thouless line, one could thus at best expect a regime of correlated spin clusters, whose size diverges as $\Gamma\to 0$. This might explain the flat maximum in the non-linear susceptibility as a function of $\Gamma$, and its decrease with $T$. However, to fully understand the dynamic response of LiY$_{1-x}$Ho$_x$F$_4$, it will be crucial to include the hyper-fine coupling to the Ho nuclear spin, especially in the quantum glassy low $T$ regime~\cite{SchechterStamp}. 

Interestingly, the susceptibility at a given point ($\Gamma,T$) in the glassy regime depends strongly on the annealing protocol~\cite{Brook}. Quantum annealing results in significantly faster typical relaxation rates and  larger $\chi'(\omega)$ than  thermal annealing. Moreover, it results in a nearly $\Gamma$-independent dissipation $\chi''(\omega)$ at low $\omega$, which was interpreted to signal a critical state. The authors suggested that faster response in quantum annealed samples implies more complete relaxation than under thermal annealing. 
However, the opposite conclusion seems  equally possible: quantum annealing remains stuck closer to the marginal surface of the rugged energy landscape, leaving more fluctuating regions with small barriers and faster relaxation times. Instead the thermally annealed 
state, if the system managed to cross barriers, looks more ``aged", being entrenched in deeper and more stable valleys, and thus displaying slower response.

\subsection{Quantum Heisenberg spin glasses}

Heisenberg quantum spin glasses attract a lot of interest because of the intriguing SYK-like spectral features they might display in their paramagnetic phase, as well as above a moderate frequency scale -- provided the mean-field  and large-$M$ predictions survive for $SU(2)$ spins in 3d. A prominent example is the cuprate La$_{2-x}$Sr$_x$CuO$_4$~\cite{Frachet2020}
which was recently shown to host spin glass order at low temperatures, all the way up to optimal doping, albeit with strong short-range antiferromagnetic correlations.
If critical spectral features as predicted for fully frustrated mean-field Heisenberg glasses also survive in such real materials, 
they might possibly be at the root of the linear temperature dependence of the resistivity often observed in this type of materials~\cite{Chowdhuryetal22}. 

{In heavy fermion compounds, such as {\em e.g.}~Y$_{1-x}$U$_x$Pd$_3$,  the dilute, randomly positioned local moments of U 
couple to conduction electrons that mediate RKKY interactions. 
%whose sign rapidly oscillates with distance and thus frustration is potentially strong. 
Such materials are promising candidates 
to find genuine metallic quantum glass phases and quantum glass transitions 
to a paramagnet, where Kondo screening of the moments dominates. 
A quantum spin glass phase and interesting non-Fermi liquid behavior at higher temperatures was indeed reported for $0.2< x < 0.4$ of the above compound~\cite{Gajewski00}. 
For a more thorough review of quantum glass candidates and their magnetic properties we refer to~\cite{Mydosh15}.  
} 

\subsection{Mean-field quantum glasses in  optical cavities}
\label{sec:cavity} 

Random and long-range  interactions among cold atoms  can be generated via light-matter coupling in multi-mode cavities with  sufficiently random mode functions \cite{GopalakrishnanCavity2009,GopalakrishnanCavity2011}. We distinguish realizations with non-intinerant and itinerant atoms, respectively.

\subsubsection{Non-itinerant fermionic atoms}

Laser-trapped, immobile atoms act as local TLS (Ising spins). They are coupled by an effective long range Ising interaction $J_{ij}$ obtained from integrating out the photons of the enclosing optical cavity. 
In cavities with a large number of modes,  the long-ranged $J_{ij}$ are nearly random, 
owing to the complex structure of the mode amplitudes.  Such systems effectively realize a TFSK model, whereby the interaction may assume a non-vanishing mean $[J_{ij}]=J_0$ \cite{StrackSachdev2011}, and can be tuned via the driving parameters of the optical cavity.
%As expected, upon increasing the interaction strength with respect to  the transverse field $\Gamma$, one can drive the quantum phase transition between a quantum paramagnet and the  quantum spin glass phase. 
Upon dialling up the average coupling $J_0$ a ferromagnetic component of the spin freezing may arise as in classical analogues. Here it essentially realizes Dicke's superradiant phase. 
%%\MM{could be skipped:} 
Under off-equilibrium conditions, as introduced by the driving of the laser cavity and the leakage of photons, 
the (steady state) phase diagram remains robust, but the critical behavior is modified~\cite{Buchhold2013}.

\subsubsection{Itinerant fermionic atoms}
\label{cavitymetallicglass} 

If the atoms are able to move, 
%from site to site of the optical lattice within the laser cavity, 
an interesting interplay between amorphous charge order 
%(and associated static disorder potentials) 
and atomic delocalization  occurs. In contrast to the case of Coulomb interactions (Sec.~\ref{sec:QEG}), the interactions remain long-ranged across the glass and the localization transitions. This allows for a glassy density order to develop upon decreasing the kinetic energy of the fermions while they are still in a metallic phase. Glassy density order in turn generates increased local disorder which eventually  localizes the fermions.

The instability of the ergodic metal to glassy order and the single particle localization threshold are well separated
in a lattice model with one fermionic level per site and close to half-filling. Thus, a metallic glass intervenes between the insulator and the ergodic metal \cite{MuellerStrackSachdev2012}. 
In contrast, for more dilute filling
the transition between ergodic metal and glassy insulator becomes first order, 
%as for $d> 2$ the local instabilities of either phase occur deep inside the region where the opposite phase is still stable. %That is, through localization the insulator enhances the disorder felt by the individual fermions and thereby self-stabilizes the glassy insulating phase. While the metallic and rather homogeneously dense phase remains stable towards a glassy instability and the occurrence of disorder. 
and in an extended parameter regime the two phases can coexist. 
How one phase nucleates out of the less favorable one 
%upon a slow tuning of system parameters 
remains an interesting open question. 
 
\section{Fermion models and the Sachdev-Ye-Kitaev model}
\label{sec:SYK}

Kitaev~\cite{Kitaev} proposed to use the Majorana fermion system
\begin{equation}
\hat {\mathcal H}_{SYK} = \frac{1}{i^{q/2}} \sum_{i_i < i_2 < \dots < i_q}^N J_{i_i i_2 \dots i_q}
\hat \psi_{i_1} \hat \psi_{i_2} \dots \hat \psi_{i_q}  
\; , 
\label{eq:SYK}
\end{equation}
with  $\hat \psi_i = \hat \psi^\dag_i$, $\{ \hat \psi_i , \hat \psi_j \}  = 2\delta_{ij} $, quenched random couplings with zero mean
and $[J^2_{i_1 \dots i_q}]=(q-1)! J^2 /N^{q-1}$, and 
four fermion ($q=4$) interactions, as a toy model 
for near-extremal black holes. Being very similar to the fermionic representation of the $SU(M\to\infty)$ Heisenberg model, it has no glassy phase.
%High energy interest resides in the fact that the 
The main reason to propose the connection with gravity is that it becomes approximately conformal in the infrared (dropping time-derivatives, low-frequency limit), with a time-reparametrization symmetry which 
is broken to SL(2,R), as expected in blackhole theories  which develop a nearly AdS$_2$ 
background~\cite{Chowdhuryetal22}. A ``Schwarzian action" describes the cost of reparametrizations~\cite{MaldacenaStanford}.
Moreover, there is non-zero entropy at $T\to 0$ (taken after $N\to\infty$) and the specific heat is linear in $T$ -- thermodynamic properties
that are expected in the black hole context as well. Finally, this system is a
``maximally chaotic/perfect scrambler'' meaning that the bound $\lambda_T \leq  2\pi T/\hbar$ 
on the Lyapunov exponent (a consequence
of the fluctuation-dissipation theorem~\cite{Pappalardi}), defined from 
the exponential growth of out-\-of-\-time-\-order correlations~\cite{Shenker}, is 
saturated. 

It was soon  realized
that Kitaev's proposal is very close to the Sachdev-Ye model, 
introduced and studied with the aim of describing non-Fermi liquid behavior in condensed matter systems hosting quantum spins (see Sec.~\ref{SUM}).
The above quenched disordered fermion system (with no glassy phase), subsequently called SYK, thus became a popular 
model in high-energy theory. Potential experimental realizations were proposed in disordered graphene flakes under strong magnetic fields~\cite{Franz18,Kruchkov22}.

In order to make a connection with the formalism we described in the rest of the 
chapter, the relevant correlator in the SYK model is $Q(\tau) = - [\langle {\rm T} \hat\psi_i(\tau) \hat\psi_j(0) \rangle ]$ (the fermionic analogue of $G_B$ of Sec.~\ref{SUM}),
which satisfies an equation like Eq.~(\ref{eq:generic}) in its replica-diagonal form (no need of off-diagonal terms here)
with 
\begin{equation}
G_0^{-1} = d_\tau \; ,  \qquad\qquad
\Sigma(\tau) = J^2 Q^3(\tau)
\; . 
\end{equation}

Let us now comment on some interesting properties of the SYK model.  Time-reparametrization symmetry also emerges 
 in mean-field glass models, where sigma models for the reparametrizations 
 were phenomenologically proposed (but not derived)~\cite{ChamonCugliandolo}.
 Coupling two (or more) SYKs, popular in high-energy studies to mimic wormholes,  is similar to coupling real replicas
 in disordered systems, a procedure that has long been used in the glass literature to access properties of the 
 free-energy landscape~\cite{FranzParisi}.
 Moreover, the relation of SYK  with tensor models without quenched disorder~\cite{Klebanov}, 
 parallels the ideas of self-generated disorder put forward in the 90s to link $p$-spin models to 
 the Mode-Coupling Theory for structural glasses~\cite{Bouchaud}. In the SYK context, the trick is to use tensor models 
 such that in a perturbative expansion only 
melon diagrams for the two-time function, and ladder diagrams for the four-time function, survive.
With quenched disorder this structure occurs thanks to the average over randomness.

These results motivated numerous studies of other quantum spin glasses~\cite{Anous,Bera}, in particular those
discussed in Sec.~\ref{sec:pair-vs-muti}. Let us just mention here the analysis of chaos
in the paramagnetic and marginal glassy phases of the 
spherical $p$-spin models~\cite{Bera}. 
%One of the aims of these papers was to study the dependence of $\lambda_T$ on the 
%strength of the quantum fluctuations and attain the classical limit. 
Quantum fluctuations were found to 
make the paramagnetic phase less and the spin glass phase more chaotic. In the classical limit $\hbar\to 0$, a
crossover from strong to weak chaos, as marked by a maximum in $\lambda_T$, arises well above $T_d$, concomitant with the onset of two-step slow relaxation.

\section{Conclusion and outlook}

This short review on quantum glassy systems focuses on the discussion of their equilibrium 
properties. We have tried to tie connections between the different systems and models presented.
On the analytic side, we mostly discussed results obtained in different mean-field limits, 
where the replica method and replica symmetry breaking schemes can be safely applied. 
The phenomenology of these disordered and glassy mean-field  models is very rich and has a wide scope.
Indeed, many applications beyond physics have been exploited, notably to optimization problems, 
but more recently fruitful connections with unexpected fields, like gravity, have been uncovered, too.
As usual, finite-dimensional physically relevant problems are not amenable to an exact treatment and much of 
what is known about them is either phenomenological or numerical.

Putting this summary together we came across a number of open issues which would be interesting
to study. We just mention three of them. Revisiting the quantum glass transitions with deconfining spinons, especially in insulating Heisenberg systems,
should be within analytic reach. We have only superficially discussed the interplay 
between glassiness and localization. There are certainly many interesting open questions there:
how and when does glassiness set in? How does a localized quantum glass melt into a metal? 
Finally, it would be welcome to establish deeper connections between SYK insights (such as the treatment of the reparametrization invariance)
and glass physics.

\bibliographystyle{ws-rv-van}
\bibliography{referencesMM}

\end{document}